\documentclass[twocolumn,showpacs,prl,floatfix,letter]{revtex4-1}
\ifx\pdfoutput\undefined
\usepackage{graphicx}
\else
\usepackage{epstopdf}
\usepackage{epsfig}
\usepackage{amssymb,amsmath,stmaryrd,tabularx}
\usepackage{wrapfig}
\usepackage{bm}
\usepackage{ulem}
\usepackage{xcolor}
\fi

\usepackage[center]{subfigure}

\begin{document}
\title{Three-body Interactions Drive the Transition to Polar Order in a Simple Flocking Model}
\author{Purba Chatterjee and Nigel Goldenfeld}
\affiliation{
Department of Physics, University of Illinois at
Urbana-Champaign, Loomis Laboratory of Physics, 1110 West Green
Street, Urbana, Illinois, 61801-3080
}


\pacs{}

\begin{abstract}
A large class of mesoscopic or macroscopic flocking theories are coarse
grained from microscopic models that feature binary interactions as the
chief aligning mechanism. However while such theories seemingly predict
the existence of polar order with just binary interactions, actomyosin
motility assay experiments show that binary interactions are
insufficient to obtain polar order, especially at high densities. To
resolve this paradox, here we introduce a solvable one-dimensional
flocking model and derive its stochastic hydrodynamics. We show that
two-body interactions are insufficient to generate polar order unless
the noise is non-Gaussian.  We show that noisy three-body interactions
in the microscopic theory allow us to capture all essential dynamical
features of the flocking transition, in systems that achieve
orientational order above a critical density.
\end{abstract}
\maketitle

The flocking transition in active matter
has been widely studied both in experiments and with theoretical models
\cite{Marchetti2013, Vicsek1995, VolkerSchaller2010, A.Ahmadi2005}. The
phase-diagram for a simple flocking theory, with alignment interactions
that tend to promote polar order is now more or less well understood
\cite{Vicsek1995, Gregoire2004, Solon2015}. What remains to be
established is what microscopic interactions give rise to the ordering
transition. On one hand, microscopic models with just binary alignment
interactions have apparently been shown to have polar ordered phases
\cite{A.Ahmadi2005,Bertin2006,Hanke2013}. On the other hand, recent
experiments on actomyosin motility assays \cite{Suzuki2015} showed
conclusively that microscopic binary interactions are insufficient to
drive the transition to polar order. In fact, at the high densities at
which the flocking transition takes place in these systems, binary
collisions constitute a very small fraction of the whole range of
interactions.

This contradiction between theoretical and experimental results prompts
us to look more closely at two-body interactions as a sufficient
mechanism for flocking.  Specifically, we examine what types of
approximation are used in theoretical models, and how they influence
the result.  Hydrodynamic theories coarse-grained from microscopic
binary collision models \cite{A.Ahmadi2005,Bertin2006,Hanke2013}
necessarily have effective higher order interactions because of the
averages taken over many two-body interactions. They therefore exhibit
different phase-diagrams from the original microscopic theory
calculated without approximations. The polar order observed in all
these cases \cite{Bertin2006,A.Ahmadi2005,Bertin2006,Hanke2013} is then
simply an artifact of uncontrolled approximations made on the
microscopic model, and is not driven by two-body interactions. In
\cite{Bertin2006}, direct simulations of a microscopic binary collision
model seemingly generate polar order, in contradiction to experimental 
results \cite{Suzuki2015}. We will show below that this result arises 
from the use of non-Gaussian noise in the simulation,
which introduces effective multi-body interactions, and changes the
phase-diagram from the case with standard Gaussian noise. We ask then,
in a model where effective interactions are not introduced by
coarse-graining or skewed noise statistics, are two-body interactions
sufficient for establishing polar order?

In this Letter, we answer this question in the negative by using a
solvable $1$D individual level flocking model which features both
two-body and three-body interactions as aligning mechanisms. Through
exact simulations of the individual level interactions, as well as
analytical calculations on the coarse-grained hydrodynamics, we show
that while two-body interactions cannot generate polar order in
agreement with experiments, stochastic three-body interactions in 
the microscopic theory generate polar order and recapitulate the 
generic dynamical features of the flocking transition.



\textit{The model:-} Our starting point is the Active Ising model (AIM) \cite{Solon2013,Solon2015a}, with modifications to highlight the roles of two-body and three-body interactions. We consider $N$ particles on a $1$D lattice of size $L$, each carrying spin $s=\pm 1$. There is no exclusion principle in this model, which allows for an arbitrary number of particles on each lattice site. Let us denote by $n_i^{\pm}$ the number of $\pm$ spins on lattice site $i$. The local densities are given by $\rho_i=n_i^+ + n_i^-$, and the local polarization/magnetization by $m=n_i^+ - n_i^-$. Self-propulsion is modeled by giving positive spins a higher probability of hopping forward than backward, and negative spins a higher probability of hopping backward than forward:
\begin{align}
N_i^+ \xrightarrow{D(1+\epsilon)} N_{i+1}^+ \hspace{10pt}&,\hspace{10pt} N_i^+ \xrightarrow{D(1-\epsilon)} N_{i-1}^+,\label{eq1}\\
N_i^- \xrightarrow{D(1-\epsilon)} N_{i+1}^- \hspace{10pt}&,\hspace{10pt} N_i^- \xrightarrow{D(1+\epsilon)} N_{i-1}^-,\label{eq2}
\end{align}
where $N_i^{\pm}$ is the population of $\pm$ spins at site $i$, $D$ is the diffusion coefficient and $\epsilon \in[0,1]$ is a measure of the self-propulsion velocity. 
The particles on a site interact with each other and flip their spin according to the following stochastic processes:
\begin{align}
N_i^- \xrightarrow{T} N_i^+ \hspace{10pt}&,\hspace{10pt} N_i^+ \xrightarrow{T} N_i^-,\label{eq3}\\
N_i^+ + N_i^-\xrightarrow{\hat{r}_2} 2N_i^+ \hspace{10pt}&,\hspace{10pt} N_i^+ + N_i^- \xrightarrow{\hat{r}_2} 2N_i^-,\label{eq4}\\
2N_i^+ + N_i^- \xrightarrow{\hat{r}_3} 3N_i^+ \hspace{10pt}&,\hspace{10pt}N_i^+ + 2N_i^- \xrightarrow{\hat{r}_3} 3N_i^-.\label{eq5}
\end{align}
The first process is a random spin flip at rate $T$, which sets the temperature in this model. The second and third processes represent two and three-body interactions respectively, and proceed at rates $\hat{r_2}$ and $\hat{r_3}$ respectively. We rescale the rates
$\hat{r}_a=r_a/\rho_i^{a-1}$ with $a=2,3$, to ensure that they remain bounded.

\begin{figure}
\begin{center}
\includegraphics[scale=0.15]{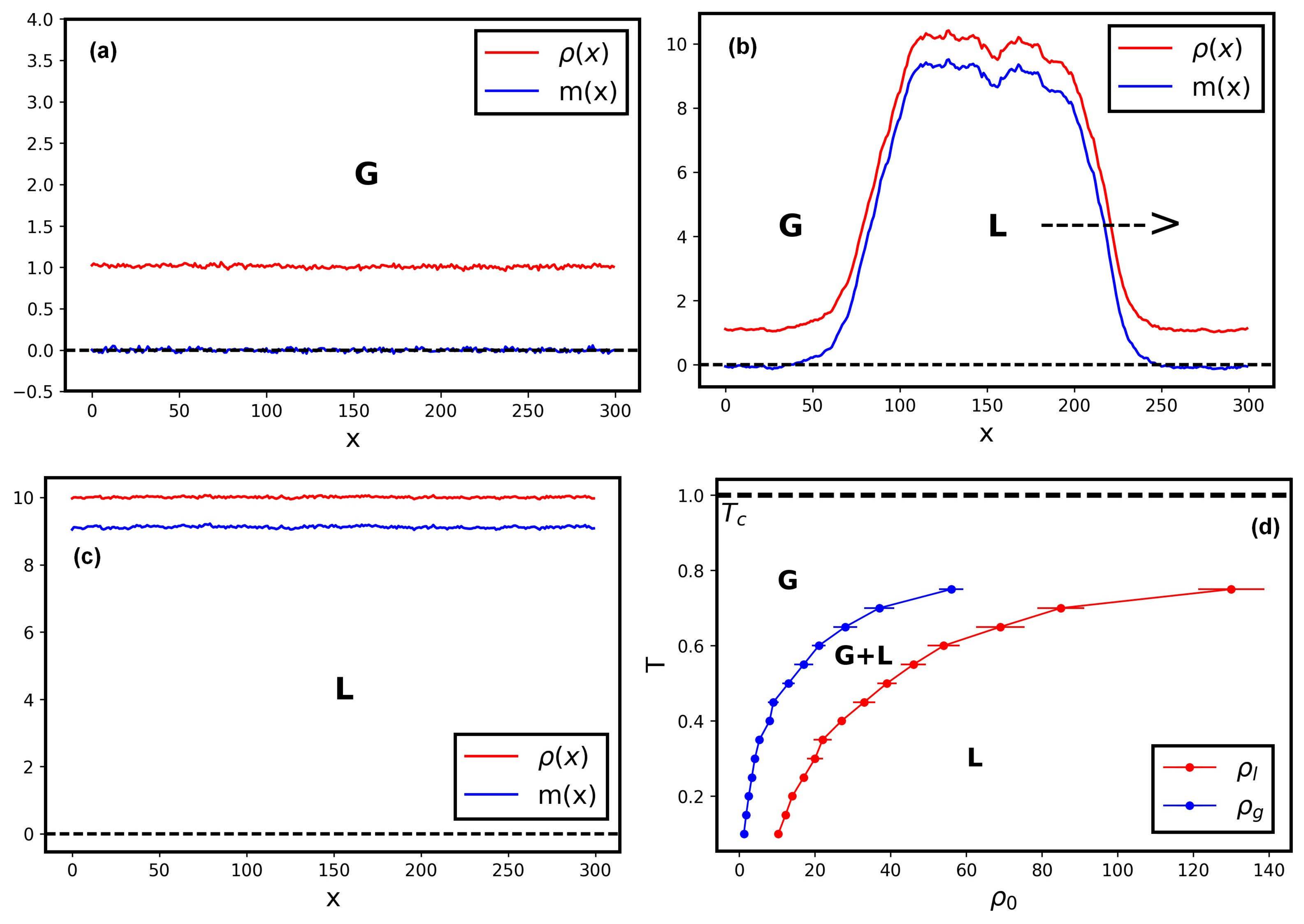}
\end{center}
\caption {\small{(Color online) Time averages of density and magnetization profiles from exact simulations of the microscopic model. (a) Disordered gas, $T=0.1, \rho_0=1.0 $. (b)  Liquid-gas coexistence, $T=0.1, \rho_0=5.0$. (c) Polar liquid, $T=0.1, \rho_0=10.0$. (d) Numerical phase-diagram: $\rho_g$ and $\rho_l$ delimit the region of existence of phase-separated profiles. $D=1, \epsilon=0.9, r_2=1, r_3=4, L=300$ for all figures.}}
    \label{fig1}
\end{figure}

\begin{figure}
\begin{center}
\includegraphics[scale=0.15]{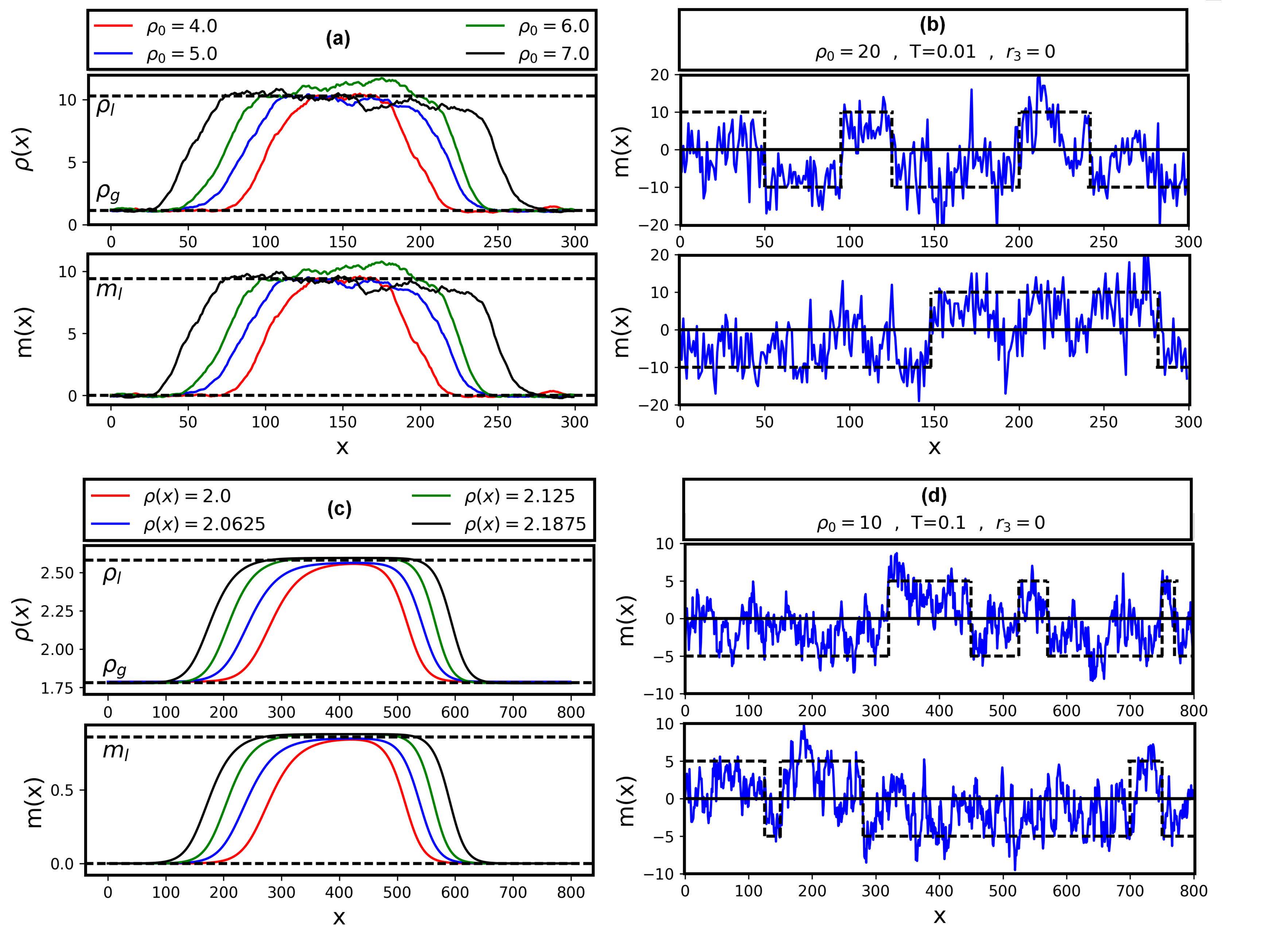}
\end{center}
\caption {\small{(Color online) \textit{Top}: Results from stochastic simulations, $D=1, \epsilon=0.9, r_2=1, L=300 $ for all figures. (a)
Phase separated profiles as a function of density from exact simulations, $T=0.1, r_3=4$. (b) Local switching of magnetization for
$r_3=0$ from exact simulations, $T=0.01, \rho_0=20.0$. \textit{Bottom}: Results from analytical model, $D=r=v=1, r_2=1, L=800$ for all figures. (c) Phase separated profiles as a function of density from WFT, $T=0.5, r_3=4$. (d) Local switching of magnetization for $r_3=0$ from simulations of the full sPDE, $T=0.1, \rho_0=10.0$.}}
    \label{fig2}
\end{figure}

\textit{Simulation results:-} We simulate the microscopic flocking model exactly using the Gillespie algorithm \cite{Gillespie1977}, and periodic boundary conditions. We set $r_3=4$ and $r_2=1$ for all calculations, both numerical and analytical, for the remainder of this letter. Varying $T$ and $\rho_0=N/L$, three distinct phases are observed. For low densities and high temperature we obtain a homogeneous disordered phase (gas), with $\langle m \rangle=0$ (Fig. \ref{fig1}(a)). For high densities and low noise a homogeneous ordered phase (liquid) is observed with $\langle m \rangle \neq 0$ (Fig. \ref{fig1}(c)). For intermediate densities $\rho_0 \in (\rho_g(T),\rho_l(T))$, we get phase-coexistence - a band of high density ordered liquid traveling in a dilute disordered gaseous background (Fig. \ref{fig1}(b)). The numerical phase-diagram of the model is shown in Fig. \ref{fig1}(d), where we plot the two coexistence lines $\rho_g$ and $\rho_l$ that delimit the existence of phase-separated profiles. Within the coexistence regime, increasing the average density $\rho_0$ at a given temperature only enlarges the liquid fraction (Fig. \ref{fig2}(a)), while keeping the density of the liquid and gas fractions constant.

Setting $r_3=0$ in the simulation results in a new phase for low densities, where we see local switching behavior of the magnetization,
resulting in short lived localized states of non-zero magnetization that do not form traveling fronts (Fig. \ref{fig2}(b)). We will show below that this behavior is noise-induced and arises solely due to the stochasticity of the binary interactions. However no homogeneous ordered phase is observed in the absence of three-body interactions.

\textit{Stochastic Hydrodynamics:-} The coupled stochastic partial differential equations (sPDE) that govern this system are given by
\begin{align}
\partial_t \rho&=D\Delta\rho-v \partial_x m,\label{eq7}\\
\partial_t m&=D \Delta m-v\partial_x\rho-m\Bigg[2\Big(T-\frac{r_3}{4}\Big)+\frac{r_3}{2}\frac{m^2}{\rho^2}\Bigg]\nonumber\\
&+2\sqrt{\frac{\beta}{\rho}\Bigg(\frac{T+\beta}{\beta}\rho^2-m^2\Bigg)}\eta,\label{eq8}
\end{align}
where $v=2D\epsilon$, $\beta=(r_2/2)+(r_3/4)$ and $\eta(x,t)$ is a Gaussian white noise that satisfies $\langle \eta(x,t)\eta(y,t')\rangle=\delta(y-x)\delta(t-t')$.
The derivation of this sPDE is given in the Supplementary Material. 
The dynamics is controlled by a multiplicative noise in $m$, whose strength varies depending upon the local magnetization and density. The rate $r_2$ does not appear in the relaxation of $m$ because the two-body interactions are symmetric, but does appear in the multiplicative noise, suggesting the possibility of a noise induced symmetry breaking via two-body interactions in the coarse-grained theory. In order to determine if this coarse-grained sPDE generates a different phase-diagram than the microscopic model, we need to study the steady states of the hydrodynamic theory.

\textit{Mean Field Theory:-} We first look at a simple mean field theory (MFT), where both fluctuations and correlations in
$m$ and $\rho$ are neglected. The working PDE in the mean-field limit is:
\begin{align}
\partial_t \rho&=D\Delta\rho-v \partial_x m,\label{eq10}\\
\partial_t m&=D \Delta m-v\partial_x\rho-m\Bigg[2\Big(T-\frac{r_3}{4}\Big)+\frac{r_3}{2}\frac{m^2}{\rho^2}\Bigg].\label{eq11}
\end{align}
For $T>r_3/4$, the only stable steady state solution is $\rho=\rho_0=N/L$, $m=m_0=0$. For $T<r_3/4$, two homogeneous ordered states become available and are linearly stable, $\rho=\rho_0=N/L$, $m=m_0=\pm \rho_0\sqrt{\frac{r_3-4T}{r_3}}$. Thus in this mean field approximation, by reducing $T$ below the critical temperature $T_c=r_3/4$, we go \textit{continuously} from a stable homogeneous disordered to a stable homogeneous ordered state, as can be seen from Fig. \ref{fig3}(a,b).\\
\begin{figure}
\begin{center}
\includegraphics[scale=0.15]{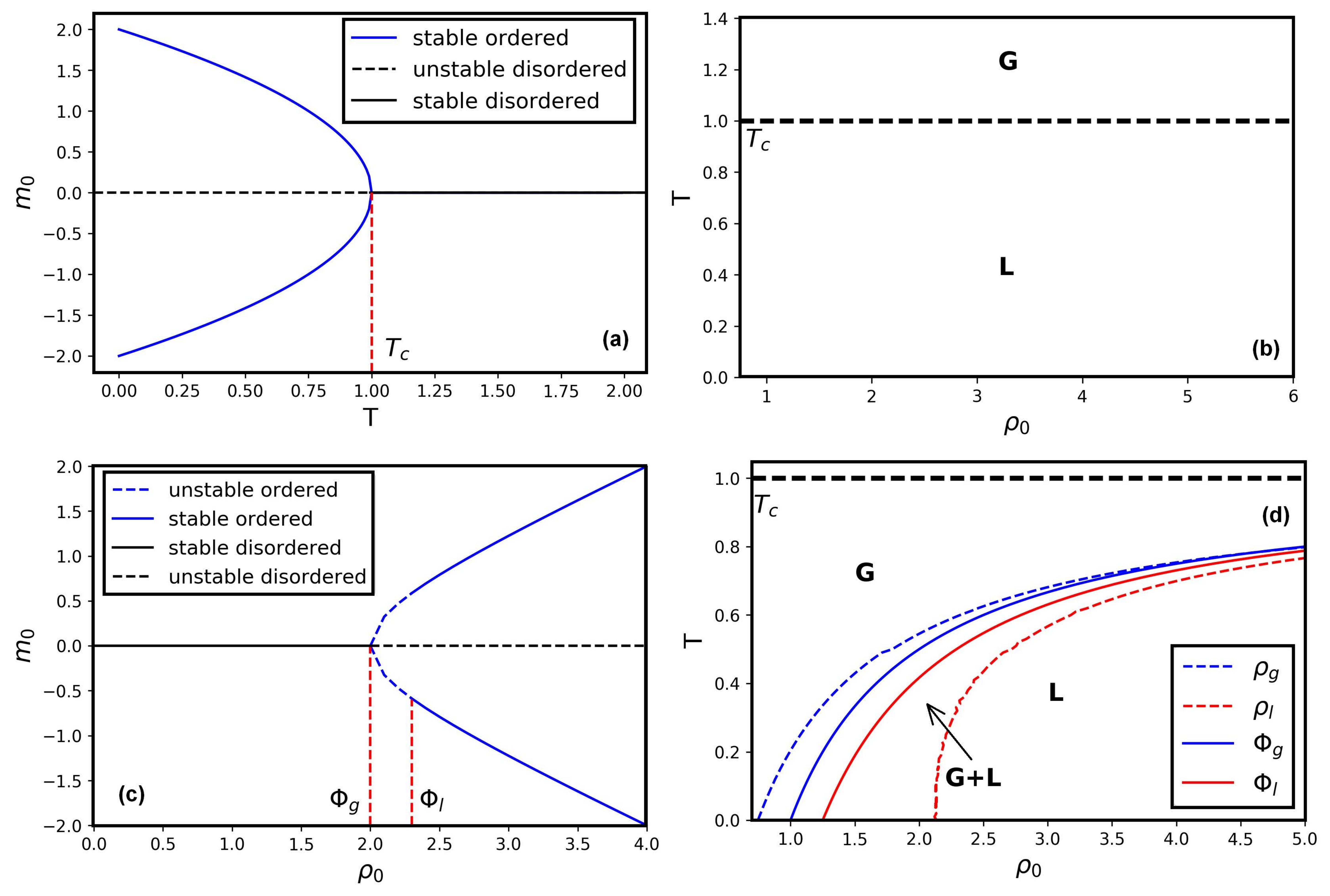}
\label{fig3}
\end{center}
\caption {\small (Color online){ \textit{Top}: phase-diagrams for the MFT with $r_3=4, r_2=1, L=800$. (a) $m_0$ vs $T$ for $\rho_0=2$. (b)
phase-diagram in $T-\rho_0$ space for the MFT. \textit{Bottom}: phase-diagrams for the WFT for $r_3=4, r_2=1, r=v=D=1,L=800$. (c) $m_0$ vs $\rho_0$ for $T=0.5$. The homogeneous ordered phase is unstable for $\rho_0 \in
(\phi_g,\phi_l)$. (d) phase-diagram in $T-\rho_0$ space for the WFT. $\phi_g$ and $\phi_l$ mark the limit of stability of the homogeneous
disordered and ordered phases respectively, for T below $T_c=r_3/4=1$. $\rho_g$ and $\rho_l$ are coexistence lines that delimit the region of existence of phase-separated profiles.}}
    \label{fig3}
\end{figure}

If $r_3=0$ , i.e, if two-body interactions are the \textit{sole} alignment mechanism in the system, the  only linearly stable homogeneous
steady state in the MFT is a disordered one with $\rho=\rho_0$ and $m=m_0=0$. This leads us to the conclusion that three-body interactions are necessary for to obtain polar order \cite{Suzuki2015}.

No phase-separated profiles are observed in this mean field limit. In the $T - \rho$ phase space this transition is depicted by a continuous line at $T=T_c$ (Fig. \ref{fig3}(b)). This is in contrast to the numerical phase-diagram shown in Fig. \ref{fig1}(d) which has a phase-separated region for all $T<T_c$. Therefore, the MFT approach misses an important dynamical feature of a typical flocking system, which invariably supports phase-separated traveling profiles at intermediate densities, resulting in a \textit{discontinuous} transition from disorder to order. The reason for this is that the MFT neglect fluctuations in $m$ and $rho$.
\begin{figure}
\begin{center}
\includegraphics[scale=0.15]{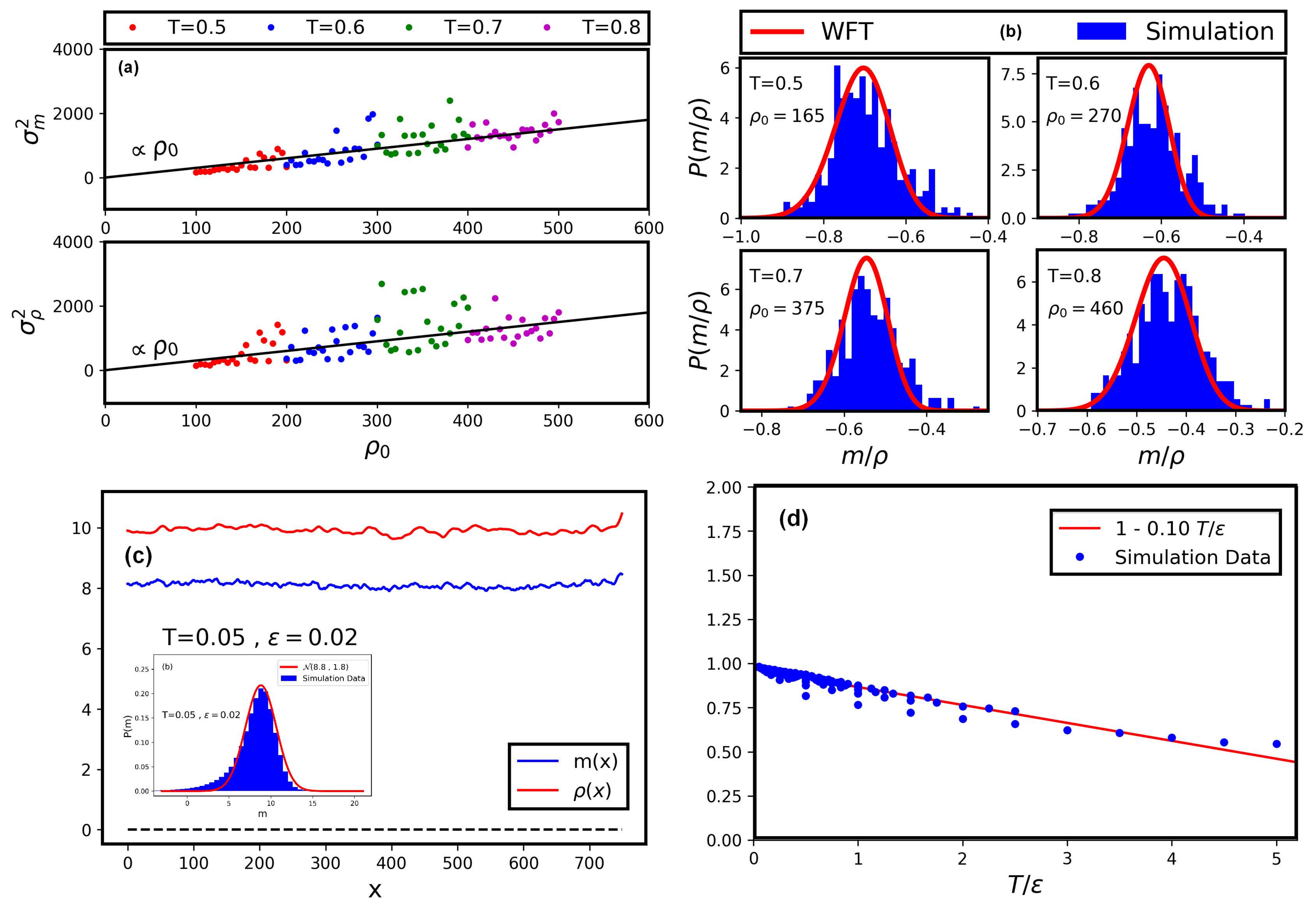}
\end{center}
\caption {\small (Color online){ \textit{Top}: Exact Simulations compared to the WFT. (a) Variances of the $\rho$ and $m$ distributions as a function of $\rho_0$. (b) Probability distribution of $m/\rho$ from Gillespie simulations (red) and as predicted by WFT (blue). \textit{Bottom}: Simulations of sPDE (\ref{eq7},\ref{eq8}) with non-Gaussian noise, $L=800, D=v=1, r_2=1$. (c) Density and magnetization profile averaged over time for $T=0.05, \epsilon=0.02, \rho_0=10.0 $. Inset: Distribution of the magnetization for the same profile, overlayed by a normal distribution with $\mu=8.8,\sigma=0.27$. (d) Plot of $\langle m\rangle ^2/\rho^2$ vs $T/\epsilon$ is found to be linear.}}
    \label{fig4}
\end{figure}

\textit{Weak Fluctuation Theory:-} We now attempt to include the effect of fluctuations generating a `weak fluctuation' approximation, following \cite{Solon2015a}. MFT assumes that the distribution of $m$ and $\rho$ as a function of space and time is given by a product of
delta functions, $P(\rho,m,x,t)=\delta(\rho(x,t)-\bar{\rho}(x,t))\delta(m(x,t)-\bar{m}(x,t))$, where $\bar{\rho}$ and $\bar{m}$ are the solutions to the mean field equations (\ref{eq10}) and (\ref{eq11}). This results in a completely deterministic time evolution of $\rho$ and $m$ given a particular set of initial conditions. The next simplest approximation is to allow $m$ and $\rho$ to have small Gaussian fluctuations about their mean-field values. 
The probability distribution of $m$ and $\rho$ is given is this case by
\begin{equation}
P(\rho,m,x,t)=\mathcal{N}(\rho-\bar{\rho},\sigma^2_{\rho})\mathcal{N}(m-\bar{m},\sigma^2_{m}),\label{eq15}
\end{equation}
where $\mathcal{N}(x-\bar{x},\sigma^2)$ is the normal distribution with mean $\bar{x}$ and variance $\sigma^2$. Since the fluctuations in $m$ and $\rho$ at $x$ are composed of $\rho(x)$ independent contributions, we expect the variances of these Gaussian distributions to be proportional to the average density, and this is supported by simulations that measure how the variances vary with respect to $\rho_0$ (Fig. \ref{fig4}(a)). We set $\sigma^2_{\rho}=a_{\rho}\bar{\rho}$ and $\sigma^2_{m}=a_{m}\bar{\rho}$, where $a_{\rho}$ and $a_m$ are
temperature dependent. 
Only the non-linear term in $m$ has to be approximated, and with (\ref{eq15}) we get
\begin{equation}
\Big{\langle} m \Big[2(T-\frac{r_3}{4}) + \frac{r_3}{2}\frac{m^2}{\rho^2}\Big]\Big{\rangle}\approx m \Big[2(T-\frac{r_3}{4}+\frac{r}{\rho}) + \frac{r_3}{2}\frac{m^2}{\rho^2}\Big],\label{eq17}
\end{equation}
where $r=3 r_3 a_m/4$, and depends only on the rate of three-body interactions $r_3$. The transition temperature is thus renormalized, and now has a density dependence:
\begin{equation}
T'_c=\frac{r_3}{4}-\frac{r}{\rho}=T^{MF}_c-\frac{r}{\rho}.\label{eq18}
\end{equation}
The Weak Fluctuation Theory (WFT) is given by:
\begin{align}
\partial_t \rho&=D\Delta\rho-v \partial_x m,\label{eq19}\\
\partial_t m&=D \Delta m-v\partial_x\rho-m\Bigg[2\Big(T-\frac{r_3}{4}+\frac{r}{\rho}\Big)+\frac{r_3}{2}\frac{m^2}{\rho^2}\Bigg].\label{eq20}
\end{align}
We will now analyze the linear stability of homogeneous steady states allowed in the WFT. For $T>T^{MF}_c=r_3/4$, the only linearly stable homogeneous steady state is disordered with $\rho_0=N/L$, $m_0=0$. For $T<r_3/4$ the homogeneous disordered state $m_0=0$ is linearly stable for all $\rho_0<\phi_g(T)$, where
\begin{equation}
\phi_g(T)=\frac{4r}{r_3-4T}.\label{eq22}
\end{equation}
For $T<r_3/4$ the homogeneous ordered state $\rho_0=N/L$, $m_0=\pm \rho_0\sqrt{\frac{(r_3-4T-4(r/\rho))}{r_3}}$, exists for all $\rho_0>\phi_g(T)$, but is linearly stable only for $\rho_0>\phi_l(T)>\phi_g(T)$, where
\begin{equation}
\phi_l=\phi_g\frac{v\sqrt{r_3[v^2 T +(D/4)(\Delta T)^2]}+2v^2T+Dr_3(\Delta T)}{4v^2T+Dr_3(\Delta T)},\label{eq24}
\end{equation}
with $\Delta T=r_3-4T$. $\phi_g$ and $\phi_l$ thus constitute the spinodal lines that mark the limit of stability of the homogeneous
disordered and ordered states respectively, and can be derived by standard linear stability analysis. Fig. \ref{fig3}(d) shows the phase-diagram in $T-\rho_0$ space for the WFT. At temperature $T$ below $T^{MF}_c$ and $\rho_0\in(\phi_g(T),\phi_l(T))$ we get the
characteristic phase-separated profiles of flocking models.
The coexistence lines $\rho_g$ and $\rho_l$ demarcate the region of existence of phase-separated profiles. Like in our exact simulations (Fig. \ref{fig2}(a)), increasing $\rho_0$ at constant $T$ within the coexistence region, simply widens the liquid domain (Fig. \ref{fig2}(c)).
The $m_0$ vs $\rho_0$ phase-diagram in WFT is shown in Fig. \ref{fig3}(c). Comparing the WFT phase-diagrams Fig. \ref{fig3}(c) and (d), to the MFT phase-diagrams Fig. \ref{fig3}(a) and (b), we see that the WFT captures the full dynamics of the microscopic theory.  

Setting $r_3=0$ has the same effect in the WFT as in the MFT- with only binary interactions, it is not possible to obtain polar order.  However, the WFT is only an approximation to the full stochastic theory, and if we consider the full sPDE (\ref{eq8}) in the absence of three-body interactions, we notice that the multiplicative noise has maximum strength at the deterministic fixed point $m(x)=0$,
and the system is thus driven away from the disordered state stochastically \cite{Biancalani2014}. This noise induced growth in
magnetization is local, giving rise to intermittent localized states, but not traveling fronts, as can be seen from simulation results (Fig. \ref{fig2}(d)) of the complete sPDE (\ref{eq8}). These localized states are analogous to the ones reported in Fig. \ref{fig2}(b) for the stochastic simulation of the microscopic theory, in the absence of three-body interactions. However, the local magnetization cannot grow without bound; when $m(x)\approx \pm m_{max}=\rho(x)\sqrt{\frac{T+(r_2/2)}{(r_2/2)}}$, the noise is at its minimum,
and the system is attracted back to the deterministic fixed point. What is observed then is a local switching behavior between $m=\pm m_{max}$. This noise-induced switching phase, with non-zero net magnetization is observed only below a critical density $\rho_c=r_2/T$ \cite{Biancalani2014}, and has possibly been identified in experiments \cite{Morris2019}.

However no global ordered phase is observed with $r_3=0$. This confirms that our hydrodynamic theory has the same qualitative phase-diagram as our microscopic model, and no artifacts introduced by coarse-graining. Thus we conclude that noisy three-body interactions are necessary and sufficient to capture the complete phase-diagram for the flocking transition.

\textit{Hydrodynamics with non-Gaussian noise:-} Seemingly at odds with our conclusion, \cite{Bertin2006} reports existence of polar order with just two-body interactions, but with non-Gausssian noise. To demonstrate that skewed noise statistics can generate artifactual polar order \cite{Bertin2006} with just two-body interactions, (see Supplementary Material for more details), we simulated the full stochastic hydrodynamics (\ref{eq7},\ref{eq8}) in the absence of three-body interactions ($r_3=0$), but with $\eta$ described by a non-Gaussian distribution, $P(\eta)=\mathcal{N}\exp\Big(-\frac{\eta^2}{2}+\epsilon(\eta^3-\eta^4/2)\Big)$. Here $\epsilon$ is a small parameter that introduces deviations from Gaussian noise. Fig. \ref{fig4}(c) shows a representative time-averaged magnetization profile, and it is clearly the polar ordered liquid phase with $\langle m \rangle \neq 0$, similar to Fig. \ref{fig1}(c). This is further confirmed by the probability distribution of magnetization for the same profile given in the inset of this figure, which peaks at a non-zero value. Fig. \ref{fig4}(d) shows a plot of $\langle m\rangle ^2/\rho^2$ vs $T/\epsilon$, which is linear. The mean magnetization satisfies:
\begin{equation}\label{eq25}
\langle m\rangle\Bigg[\Big(T-\frac{\epsilon r_3^{eff}}{4}\Big)+\frac{\epsilon r_3^{eff}}{2}\frac{\langle m \rangle^2}{\rho^2}\Bigg]=0,
\end{equation}
where $r_3^{eff}$ is calculated from the slope and intercept in Fig. \ref{fig4}(d). This is identical to the relaxation term in our full sPDE with three-body interactions (\ref{eq8}). Evidently, the non-Gaussian part of the noise statistics introduces ``effective" three-body interactions even when they are not explicitly present in the microscopic theory, allowing us to observe polar ordered phases. Setting $\epsilon=0$ makes the noise purely Gaussian and consequently we do not observe polar order anymore, resolving the contradiction with \cite{Bertin2006}.

\textit{Discussion}: Despite not being explicitly considered, correlations between $m$ and $\rho$ are still preserved in the WFT by approximating fluctuations in $m$ to have variances proportional to the average density. From the WFT (\ref{eq15}), a closed form expression can be derived for the probability distribution of $m/\rho$, which follows a shifted Cauchy distribution (see Supplementary Material). Fig. \ref{fig4}(b) shows a comparison between the probability distribution
of $m/\rho$ from exact simulations (blue histogram) and from WFT predictions (red curve) for four different values of noise and we can
see that the agreement is satisfactory.


Note that because in our minimal 1D model the orientations are discrete, the two-body interaction (\ref{eq4}) results in one particle following the other similar to the $\omega=0,1$ case in \cite{Thuroff2013}, for which the authors observed no ordered phase. However, even with continuous orientations and in $2$D, direct simulations of microscopic binary interactions in hard rods and stiff polymers show no ordering transition \cite{Thuroff2013}. Coarse-grained hydrodynamic theories from binary collision models have been shown to have ordered phases \cite{A.Ahmadi2005,Bertin2006,Hanke2013}, but this is because coarse-graining involves taking averages over pre-collision angles and impact parameters for many individual interactions and introduces effective many body interactions in the process. All such hydrodynamic theories have a relaxation term for the momentum order parameter, in our case $m$, of the form $-m(a+bm^2)$. The cubic term in $m$, essential for the existence of the ordered phase when $a<0$, represents an effective, if not explicit three-body interaction in the microscopic theory. Thus we expect our conclusion to hold in higher dimensions and with continuous orientations. It is important to note that the hydrodynamic theory that is obtained by coarse-graining the $1$D AIM with just binary interactions would not have such effective three-body terms simply because these binary interactions always occur with the same orientation, thus precluding the need to average over such orientations.

Finally, we checked that including four-body or higher interactions only renormalizes the transition temperature without qualitatively changing the three-body phase-diagram.

We thank M. Cristina Marchetti for useful discussions. We also thank the referees for helpful comments that improved the presentation of this work.

\bibliographystyle{apsrev4-1}
\bibliography{Purba_flocking_threebody}

\pagebreak
\widetext
\begin{center}
\textbf{\large Supplementary Material for:\\ \vspace{5pt}Three-body Interactions Drive the Transition to Polar Order in a Simple Flocking Model\\}
\vspace{5pt}
Purba Chatterjee and Nigel Goldenfeld
\end{center}
\setcounter{equation}{0}
\setcounter{figure}{0}
\setcounter{table}{0}
\setcounter{page}{1}
\makeatletter
\renewcommand{\theequation}{S\arabic{equation}}
\renewcommand{\thefigure}{S\arabic{figure}}
\renewcommand{\bibnumfmt}[1]{[S#1]}
\renewcommand{\citenumfont}[1]{S#1}

This Supplementary Material presents three calculations referred to in the main text. The first shows that two-body interactions in the presence of non-Gaussian noise statistics can give rise to polar order, as opposed to when the noise is standard Gaussian. This clarifies why in [$7$] of the main text, the authors observe ordered phases with just two-body interactions, because the noise used is non-Gaussian. The second calculation is the derivation of the stochastic hydrodynamics, given in Eq. ($6$) and ($7$) in the main paper. The third is the calculation of the probability distribution for $m/\rho$ from the Weak Fluctuation Theory (WFT), given in Eq. ($10$) in the main paper.
\section{ I. Non-Gaussian noise statistics}
In this section, we discuss simulation results of the full sPDE (Eq. ($6$) and ($7$) in the main text), in the absence of three-body interactions ($r_3=0$), and with $\eta$ described by a non-Gaussian distribution. The working sPDE in this case is
\begin{align}
\partial_t \rho&=D\Delta\rho-v \partial_x m,\label{eq1.1}\\
\partial_t m&=D \Delta m-v\partial_x\rho-2mT+\sqrt{\frac{2r_2}{\rho}\Bigg(\frac{2T+r_2}{r_2}\rho^2-m^2\Bigg)}\eta,\label{eq1.2}
\end{align}
where
\begin{equation}\label{eq1.3}
P(\eta)=\mathcal{N}\exp\Big(-\frac{\eta^2}{2}+\epsilon(\eta^3-\eta^4/2)\Big).
\end{equation}

\begin{figure}
\begin{center}
\includegraphics[scale=0.4]{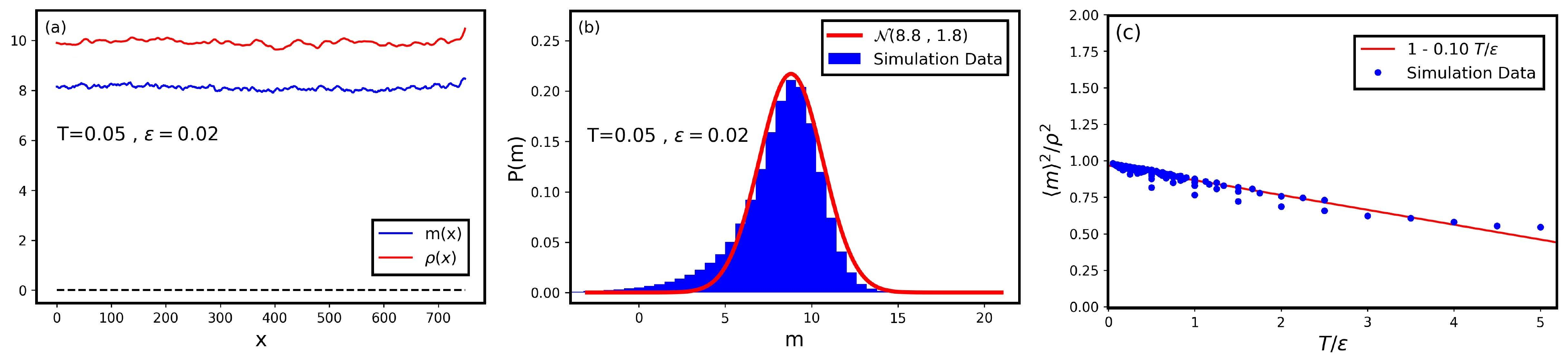}

\end{center}
\caption {\small{(Color online) (a) Density and magnetization profile averaged over time for $T=0.05, \epsilon=0.02, \rho_0=10.0 $. (b)  Distribution of the magnetization for the profile shown in (a), overlayed by a normal distribution with $\mu=8.8,\sigma=0.27$. (c) Plot of $\langle m\rangle ^2/\rho^2$ vs $T/\epsilon$ is found to be linear. $D=1, r_2=1, L=800$ for all figures.}}
    \label{fig5}
\end{figure}

Here $\epsilon$ is a small parameter that introduces deviations from Gaussian noise statistics. We simulated this sPDE for different values of $T$ and $\epsilon$ in order to determine the steady state solutions. Fig. \ref{fig5} shows results from these simulations for $L=800$ and $D=v=1$. Fig. \ref{fig5}(a) shows a representative magnetization profile averaged over time, and we immediately see that this is the homogeneous polar ordered liquid phase with $\langle m \rangle \neq 0$ similar to Fig. $1$(c) in the main text. Fig. \ref{fig5}(b) shows the probability distribution of the magnetization for the same profile, and we find that it can be approximated reasonably well with a normal distribution $\mathcal{N}(\mu,\sigma)$, with mean $\mu=8.8$ and standard deviation $\sigma=1.8$. Fig. \ref{fig5}(c) shows a plot of $\langle m\rangle ^2/\rho^2$ vs $T/\epsilon$, which is linear. The mean magnetization satisfies:
\begin{equation}\label{eq1.4}
\langle m\rangle\Bigg[\Big(T-\frac{\epsilon r_3^{eff}}{4}\Big)+\frac{\epsilon r_3^{eff}}{2}\frac{\langle m \rangle^2}{\rho^2}\Bigg]=0,
\end{equation}
where $r_3^{eff}$ is calculated from the slope and intercept in Fig. \ref{fig5}(c). Thus the effective relaxation term for magnetization is modified by the non-Gaussian part of the noise from $-2mT$ in Eq. (\ref{eq1.2}) to $-m\Big[\big(T-\frac{\epsilon r_3^{eff}}{4}\big)+\frac{\epsilon r_3}{2}\frac{m^2}{\rho^2}\Big]$. This is identical to the relaxation term in our full sPDE with three-body interactions (Eq. ($7$) in the main text). This leads us to the conclusion that the non-Gaussian part of the noise statistics introduces ``effective" three-body interactions even when they are not explicitly present in the microscopic theory, allowing us to observe polar ordered phases. Setting $\epsilon=0$ returns the relaxation term to $-2mT$ and makes the noise purely Gaussian - consequently we do not observe polar order anymore, as reported in the main text. This resolves the contradiction with [$7$] which reports polar ordered phases with just two-body interactions but with non-Gaussian noise. 
To see how such higher order terms are generated by non-Gaussian noise statistics, let us consider a simple stochastic Ordinary Differential Equation (sODE) of the form
\begin{equation}\label{eq1.5}
\dot{x}= \frac{1}{\tau}f_0(x(t))+\frac{1}{\tau}\eta(t),
\end{equation}
where $f_0(x(t))$ is a force that depends on $x(t)$, $\tau$ is a relevant time-scale, and
\begin{equation}\label{eq1.6}
P(\eta)=\mathcal{N}\exp\Big(-\frac{\eta^2}{2\sigma_0^2}+\epsilon(a\eta^3-b\eta^4)\Big),
\end{equation}
making the noise statistics non-Gaussian for non-zero $\epsilon$. The path-integral representation for the transition probability, corresponding to Eq. (\ref{eq1.5}) is
\begin{equation}\label{eq1.7}
P(x_b,t_b|x_a,t_a)=\int_{x(t_a)=x_a}^{x(t_b)=x_b} \mathcal{D}[x(t)]\hspace{5pt}\mathcal{D}[\eta(t)]\hspace{5pt}P(\eta)\delta\big(\tau\dot{x}_{\eta}-f_0(x_{\eta})-\eta\big),
\end{equation}
where in the delta function we use the subscript $\eta$ on $x$ to refer to a particular solution of $x(t)$ for a given realization of $\eta$. Performing the $\eta$ integral is trivial and upto integration constants, we obtain
\begin{equation}\label{eq1.8}
P(x_b,t_b|x_a,t_a)=\int_{x(t_a)=x_a}^{x(t_b)=x_b} \mathcal{D}[x(t)]\hspace{5pt}\exp\{-S\},
\end{equation}
where the action $S$ is given by
\begin{equation}\label{eq1.9}
S=\int ds\hspace{5pt}\Bigg(\frac{\big(\tau\dot{x}(s)-f_0(x(s))\big)^2}{2\sigma_0^2}-\epsilon\Big[a\big(\tau\dot{x}(s)-f_0(x(s))\big)^3-b\big(\tau\dot{x}(s)-f_0(x(s))\big)^4\Big]\Bigg).
\end{equation}
Now, we want to construct the effective Gaussian theory for this system, and so we define the effective force $f_{\epsilon}$, where
\begin{equation}\label{eq1.10}
f_{\epsilon} =f_0 + \epsilon f_1 + \epsilon^2 f_2 + \mathcal{O}(\epsilon^3),
\end{equation}
and effective standard deviation $\sigma_{\epsilon}$, where
\begin{equation}\label{eq1.11}
\frac{1}{2\sigma_{\epsilon}^2} =\frac{1}{2\sigma_0^2} + \epsilon \phi_1 + \epsilon^2 \phi_2 + \mathcal{O}(\epsilon^3).
\end{equation}
The action $S$ can be written in terms of $f_{\epsilon}$ and $\sigma_{\epsilon}$ as
\begin{equation}\label{eq1.12}
S=\int ds\hspace{5pt}\Bigg(\frac{\big(\tau\dot{x}(s)-f_{\epsilon}(x(s))\big)^2}{2\sigma_{\epsilon}^2}\Bigg).
\end{equation}
The Langevin equation or sODE corresponsing to this action is
\begin{equation}\label{eq1.13}
\dot{x}= \frac{1}{\tau}f_{\epsilon}(x(t))+\frac{1}{\tau}\xi(t),
\end{equation}
where
\begin{equation}\label{eq1.14}
P(\xi)=\frac{1}{\sqrt{2\pi\sigma_{\epsilon}^2}}e^{-\frac{\xi^2}{2\sigma_{\epsilon}^2}}.
\end{equation}
We determine $f_1, f_2, \phi_1, \phi_2$ by expanding the integrand in Eq. (\ref{eq1.12}) in orders of $\epsilon$, comparing this with the integrand in Eq. (\ref{eq1.9}), and solving order by order. In the expansion of Eq. (\ref{eq1.12}), we neglect powers of $\dot{x}$ higher than two, which is justified for $t>>\tau$. We obtain
\begin{align}
f_{\epsilon}&=f_0 -\epsilon \sigma_0^2 (a f_0^2+2bf_0^3) + \epsilon^2 \sigma_0^4 (3a^2f_0^3 + 10abf_0^4 + 8b^2f_0^5) + \mathcal{O}(\epsilon^3),\label{eq1.15}\\
\frac{1}{2\sigma_{\epsilon}^2} &=\frac{1}{2\sigma_0^2} + \epsilon (2af_0+3bf_0^2) + \frac{\epsilon^2}{2} \sigma_0^2 (a^2f_0^2 + 4abf_0^3 + 4b^2f_0^4) + \mathcal{O}(\epsilon^3).\label{eq1.16}
\end{align}
If we consider $f_0=-x$, it is clear from Eq. (\ref{eq1.15}) that the non-Gaussian part of the noise statistics generates effective higher order interactions even at first order in $\epsilon$, including an effective three-body term. Thus with non-Gaussian noise, the deterministic relaxation of the order parameter is altered, allowing for additional steady state distributions and changing the phase-diagram, as is confirmed by our simulation results (Fig. \ref{fig5}).

\section{ II. Derivation of the stochastic Hydrodynamics} 
In this section, we derive the full sPDE, Eq.(6) and (7) in the main text, starting with the $1$D AIM. Let us denote the state of the system by a $2$L dimensional vector $\mathbf{n}=\{n_1^+,n_1^-,n_2^+,n_2^-,...,n_L^+,n_L^-\}$. The stochastic processes taking place in the system are:
\begin{align}
n_i^- \xrightarrow{T} n_i^+ \hspace{10pt}&,\hspace{10pt} n_i^+ \xrightarrow{T} n_i^-,\label{eq1}\\
n_i^+ + n_i^-\xrightarrow{\hat{r_2}} 2n_i^+ \hspace{10pt}&,\hspace{10pt} n_i^+ + n_i^- \xrightarrow{\hat{r_2}} 2n_i^-,\label{eq2}\\
2n_i^+ + n_i^- \xrightarrow{\hat{r_3}} 3n_i^+ \hspace{10pt}&,\hspace{10pt}n_i^+ + 2n_i^- \xrightarrow{\hat{r_3}} 3n_i^-,\label{eq3}\\
n_i^+ \xrightarrow{D(1+\epsilon)} n_{i+1}^+ \hspace{10pt}&,\hspace{10pt} n_i^+ \xrightarrow{D(1-\epsilon)} n_{i-1}^+,\label{eq4}\\
n_i^- \xrightarrow{D(1-\epsilon)} n_{i+1}^- \hspace{10pt}&,\hspace{10pt} n_i^- \xrightarrow{D(1+\epsilon)} n_{i-1}^-.\label{eq5}
\end{align}

There are thus six processes that change the state of the system, one in which a positive spin flips, another in which a negative spin flips, two in which a positive spin hops to a neighboring site, and two more in which a negative spin hops to a neighboring site. The transition rates for these processes are given by:

\begin{align}
W_f^+(\mathbf{n})&\equiv W_{f}^+\big(\mathbf{\hat{n}},n_i^++1,n_i^--1|\mathbf{n}\big) = n_i^-\big[T+r_2 (n_i^+/\rho_i) +r_3 ((n_i^+)^2/\rho_i^2)\big],\label{eq6}\\
W_f^-(\mathbf{n})&\equiv W_{f}^-\big(\mathbf{\hat{n}},n_i^+-1,n_i^-+1|\mathbf{n}\big) = n_i^+\big[T+r_2 (n_i^-/\rho_i) +r_3 ((n_i^-)^2/\rho_i^2)\big],\label{eq7}\\
W_{h,+}^+(\mathbf{n})&\equiv W_{h,+}^+\big(\mathbf{\hat{n}},n_i^+-1,n_{i+1}^++1|\mathbf{n}\big) = D(1+\epsilon)n_i^+,\label{eq8}\\
W_{h,-}^+(\mathbf{n})&\equiv W_{h,-}^+\big(\mathbf{\hat{n}},n_i^+-1,n_{i-1}^++1|\mathbf{n}\big) = D(1-\epsilon)n_i^+,\label{eq9}\\
W_{h,+}^-(\mathbf{n})&\equiv W_{h,+}^-\big(\mathbf{\hat{n}},n_i^--1,n_{i+1}^-+1|\mathbf{n}\big) = D(1-\epsilon)n_i^-,\label{eq10}\\
W_{h,-}^-(\mathbf{n})&\equiv W_{h,-}^-\big(\mathbf{\hat{n}},n_i^--1,n_{i-1}^-+1|\mathbf{n}\big) = D(1+\epsilon)n_i^-,\label{eq11}
\end{align} 
where $\mathbf{\hat{n}}$ represents the subset of the system state that remains unchanged in that particular process. We have also rescaled the rates $\hat{r_i}=r_i/(\rho_i)^{i-1}$, $i=2,3$ so that they remain bounded. The master equation for the probability $P(\mathbf{n},t)$ is given by:
\begin{equation}
\partial_t P(\mathbf{n},t)=\sum_{\mathbf{n^{'}}}\Big{\{}W(\mathbf{n}|\mathbf{n^{'}})P(\mathbf{n^{'}},t)-W(\mathbf{n^{'}}|\mathbf{n})P(\mathbf{n},t)\Big{\}}.\label{eq12}
\end{equation}
Defining local creation and destruction operators:
\begin{equation}
a_i^{\pm}f(n_i^+)\equiv f(n_i^+\pm 1)\hspace{10pt},\hspace{10pt}b_i^{\pm}f(n_i^-)\equiv f(n_i^-\pm 1),\label{eq13}
\end{equation}
we can then write down the master equation as
\begin{align}
\partial_t P(\mathbf{n},t)&=\sum_i \Big{\{} \big(a_i^-b_i^+-1\big)W_f^+(\mathbf{n}) + \big(a_i^+b_i^--1\big)W_f^-(\mathbf{n}) \nonumber\\
&+ \big(a_i^+a_{i+1}^--1\big)W_{h,+}^+(\mathbf{n}) + \big(a_i^+a_{i-1}^--1\big)W_{h,-}^+(\mathbf{n}) + \big(b_i^+b_{i+1}^--1\big)W_{h,+}^-(\mathbf{n}) + \big(b_i^+b_{i-1}^--1\big)W_{h,-}^-(\mathbf{n})\Big{\}}P(\mathbf{n},t),\label{eq14}\\
&=\sum_i \Big{\{} \big(a_i^-b_i^+-1\big)n_i^-\big[T + r_2 (n_i^+/\rho_i) + r_3 ((n_i^+)^2/\rho_i^2)\big] \nonumber\\
&+ \big(a_i^+b_i^--1\big)n_i^+\big[T + r_2 (n_i^-/\rho_i) + r_3 ((n_i^-)^2/\rho_i^2)\big] \nonumber\\
&+ \big(a_i^+a_{i+1}^--1\big)n_i^+ D(1+\epsilon) + \big(a_i^+a_{i-1}^--1\big)n_i^+ D(1-\epsilon) \nonumber\\
&+ \big(b_i^+b_{i+1}^--1\big)n_i^- D(1-\epsilon) + \big(b_i^+b_{i-1}^--1\big)n_i^- D(1+\epsilon)\Big{\}}P(\mathbf{n},t).\label{eq15}
\end{align}
In the continuum limit, $i\to x$ and $f_i \to f(x)$. The creation and destruction operators then become
\begin{equation}
a^{\pm}(y)f[n^+(x)]\equiv f[n^+(x)\pm \Delta\delta(y-x)]\hspace{10pt},\hspace{10pt}b^{\pm}(y)f[n^-(x)]\equiv f[n^-(x)\pm \Delta\delta(y-x)],\label{eq16}
\end{equation}
and the master equation can be written in the continuum limit as
\begin{align}
\partial_t P(n^+,n^-,t)&= \frac{1}{\Delta}\int {dx\hspace{5pt} \Big(a^-(x)b^+(x)-1\Big) n^-(x)\Big[T + r_2 \big(n^+(x)/\rho(x)\big) + r_3 \big((n^+(x))^2/\rho(x)^2\big)\Big]P(n^+,n^-,t)}\nonumber\\
& + \frac{1}{\Delta}\int{dx\hspace{5pt} \Big(a^+(x)b^-(x)-1\Big)n^+(x)\Big[T + r_2 \big(n^-(x)/\rho(x)\big) + r_3 \big((n^-(x))^2/\rho(x)^2\big)\Big]P(n^+,n^-,t) }\nonumber\\
& + \frac{D(1+\epsilon)}{\Delta} \int{dx\hspace{5pt}\int{dy\hspace{5pt} \Big(a^+(x)a^-(y)-1\Big)n^+(x)\delta(y-x-a)P(n^+,n^-,t)}} \nonumber\\
& + \frac{D(1-\epsilon)}{\Delta} \int{dx\hspace{5pt}\int{dy\hspace{5pt} \Big(a^+(x)a^-(y)-1\Big)n^+(x)\delta(y-x+a)P(n^+,n^-,t)}} \nonumber\\
& + \frac{D(1-\epsilon)}{\Delta} \int{dx\hspace{5pt}\int{dy\hspace{5pt} \Big(b^+(x)b^-(y)-1\Big)n^-(x)\delta(y-x-a)P(n^+,n^-,t)}} \nonumber\\
& + \frac{D(1+\epsilon)}{\Delta} \int{dx\hspace{5pt}\int{dy\hspace{5pt} \Big(b^+(x)b^-(y)-1\Big)n^-(x)\delta(y-x+a)P(n^+,n^-,t)}},\label{eq17}
\end{align}
where a is the lattice spacing, and $\Delta=1$. This master equation can be converted into a Fokker-Planck Equation by means of a Kramers-Moyal expansion of the creation and destruction operators, truncated at second-order:
\begin{align}
a^{\pm}(x)&\approx 1\pm \Delta\frac{\delta}{\delta n^+(x)} +\frac{\Delta^2}{2}\frac{\delta^2}{\delta n^+(x)^2},\label{eq18}\\
b^{\pm}(x)&\approx 1\pm \Delta\frac{\delta}{\delta n^-(x)} +\frac{\Delta^2}{2}\frac{\delta^2}{\delta n^-(x)^2},\label{eq19}.
\end{align}

Given the large length scales of spatial variation in the hydrodynamic limit, this truncation is justified. With Eq.(\ref{eq18}) and (\ref{eq19}), the master equation can be written as:
\begin{align}
\partial_t P(n^+,n^-,t)&= \int{dx\hspace{5pt} \Big[\frac{\delta}{\delta n^+(x)}-\frac{\delta}{\delta n^-(x)}\Big]\Big[T\big(n^+(x)-n^-(x)\big)+r_3 \frac{n^+(x)n^-(x)}{\rho^2(x)}\big(n^-(x)-n^+(x)\big)\Big]P(n^+,n^-,t)},\nonumber\\
& + \int{dx\hspace{5pt} \frac{\delta}{\delta n^+(x)}\Big[2Dn^+(x)-D(1+\epsilon)n^+(x-a)-D(1-\epsilon)n^+(x+a)\Big]P(n^+,n^-,t)},\nonumber\\
& + \int{dx\hspace{5pt} \frac{\delta}{\delta n^-(x)}\Big[2Dn^-(x)-D(1-\epsilon)n^-(x-a)-D(1+\epsilon)n^-(x+a)\Big]P(n^+,n^-,t)},\nonumber\\
& + \frac{\Delta}{2} \int{dx\hspace{5pt} \Big[\frac{\delta}{\delta n^+(x)}-\frac{\delta}{\delta n^-(x)}\Big]^2\Big[T\big(n^+(x)+n^-(x)\big)+ \frac{n^+(x)n^-(x)}{\rho^2(x)}\Big(2r_2 \rho(x)+r_3\big(n^-(x)+n^+(x)\big)\Big)\Big]P(n^+,n^-,t)},\nonumber\\
& + \frac{\Delta}{2}\int{dx\hspace{5pt}\int{dy\hspace{5pt} \frac{\delta^2}{\delta n^+(x)\delta n^+(y)}\Big[2Dn^+(x)+D(1+\epsilon)n^+(x-a)+D(1-\epsilon)n^+(x+a)\Big]\delta(y-x)P(n^+,n^-,t)}},\nonumber\\
& + \frac{\Delta}{2}\int{dx\hspace{5pt}\int{dy\hspace{5pt} \frac{\delta^2}{\delta n^-(x)\delta n^-(y)}\Big[2Dn^-(x)+D(1-\epsilon)n^-(x-a)+D(1+\epsilon)n^-(x+a)\Big]\delta(y-x)P(n^+,n^-,t)}},\nonumber\\
& - \frac{\Delta}{2}\int{dx\hspace{5pt}\int{dy\hspace{5pt} \frac{\delta^2}{\delta n^+(x)\delta n^+(y)}\Big[2D(1+\epsilon)n^+(x)\delta(y-x-a)+2D(1-\epsilon)n^+(x)\delta(y-x+a)\Big]P(n^+,n^-,t)}},\nonumber\\
& - \frac{\Delta}{2}\int{dx\hspace{5pt}\int{dy\hspace{5pt} \frac{\delta^2}{\delta n^-(x)\delta n^-(y)}\Big[2D(1-\epsilon)n^-(x)\delta(y-x-a)+2D(1+\epsilon)n^-(x)\delta(y-x+a)\Big]P(n^+,n^-,t)}}.\label{eq20}
\end{align}

In the last two integrals, we can expand the $\delta(y-x\pm a)$ under the integral sign, to get:
\begin{align}
n^{\pm}(x)\delta(y-x\pm a)&\approx n^{\pm}(x)\Big{\{}\delta(y-x)\pm a \delta^{'}(y-x)+\frac{a^2}{2}\delta^{''}(y-x)\Big{\}},\nonumber\\
&=\delta(y-x)\Big{\{}n^{\pm}(x)\mp a \partial_x n^{\pm}(x)+\frac{a^2}{2}\partial_{xx}n^{\pm}(x)\Big{\}},\nonumber\\
&=n^{\pm}(x\mp a)\delta(y-x).\label{eq21}
\end{align}

Using Eq.(\ref{eq21}) the master equation becomes:

\begin{align}
\partial_t P(n^+,n^-,t)&= \int{dx\hspace{5pt} \Big[\frac{\delta}{\delta n^+(x)}-\frac{\delta}{\delta n^-(x)}\Big]\Big[T\big(n^+(x)-n^-(x)\big)+r_3 \frac{n^+(x)n^-(x)}{\rho^2(x)}\big(n^-(x)-n^+(x)\big)\Big]P(n^+,n^-,t)},\nonumber\\
& + \int{dx\hspace{5pt} \frac{\delta}{\delta n^+(x)}\Big[2Dn^+(x)-D(1+\epsilon)n^+(x-a)-D(1-\epsilon)n^+(x+a)\Big]P(n^+,n^-,t)},\nonumber\\
& + \int{dx\hspace{5pt} \frac{\delta}{\delta n^-(x)}\Big[2Dn^-(x)-D(1-\epsilon)n^-(x-a)-D(1+\epsilon)n^-(x+a)\Big]P(n^+,n^-,t)},\nonumber\\
& + \frac{\Delta}{2} \int{dx\hspace{5pt} \Big[\frac{\delta}{\delta n^+(x)}-\frac{\delta}{\delta n^-(x)}\Big]^2\Big[T\big(n^+(x)+n^-(x)\big)+ \frac{n^+(x)n^-(x)}{\rho^2(x)}\Big(2r_2 \rho(x)+r_3\big(n^-(x)+n^+(x)\big)\Big)\Big]P(n^+,n^-,t)},\nonumber\\
& + \frac{\Delta}{2}\int{dx\hspace{5pt}\int{dy\hspace{5pt} \frac{\delta^2}{\delta n^+(x)\delta n^+(y)}\Big[2Dn^+(x)-D(1-3\epsilon)n^+(x-a)-D(1+3\epsilon)n^+(x+a)\Big]\delta(y-x)P(n^+,n^-,t)}},\nonumber\\
& + \frac{\Delta}{2}\int{dx\hspace{5pt}\int{dy\hspace{5pt} \frac{\delta^2}{\delta n^-(x)\delta n^-(y)}\Big[2Dn^-(x)-D(1+3\epsilon)n^-(x-a)-D(1-3\epsilon)n^-(x+a)\Big]\delta(y-x)P(n^+,n^-,t)}}.\label{eq22}
\end{align}

We want to write down the Fokker-Planck equation for $\rho(x)=n^+(x)+n^-(x)$ and $m(x)=n^+(x)-n^-(x)$ as the variables. The derivatives then become:
\begin{equation}
\frac{\delta}{\delta n^+}=\frac{\delta}{\delta \rho}+\frac{\delta}{\delta m}\hspace{10pt},\hspace{10pt}\frac{\delta}{\delta n^-}=\frac{\delta}{\delta \rho}-\frac{\delta}{\delta m}.\label{eq23}
\end{equation}
Finally, after expanding $n^{\pm}(x\pm a)$ around x,
\begin{equation}
n^{\pm}(x\pm a)\approx n^{\pm}(x) \pm a \partial_xn^{\pm}(x)+ \frac{a^2}{2}\partial_{xx}n^{\pm}(x),\label{eq24}
\end{equation}
we obtain the FPE for the probability distribution $P(\rho,m,t)$:

\begin{align}
\partial_t P(\rho,m,t)&=-\int{dx\hspace{5pt}\Big{\{}\frac{\delta}{\delta \rho(x)}\Big[\mathcal{A}_{\rho}(\rho,m,x)P\Big]+\frac{\delta}{\delta m(x)}\Big[\mathcal{A}_m(\rho,m,x)P\Big]\Big{\}}} \nonumber \\
& + \frac{\Delta}{2} \int dx\hspace{5pt}\int dy\hspace{5pt}\Big{\{}\frac{\delta^2}{\delta \rho(x)\delta \rho(y)}\Big[\mathcal{B}_{\rho,\rho}(\rho,m,x,y)P\Big] \nonumber \\
&+ \frac{\delta^2}{\delta m(x)\delta m(y)}\Big[\mathcal{B}_{m,m}(\rho,m,x,y)P\Big]+2\frac{\delta^2}{\delta \rho(x)\delta m(y)}\Big[\mathcal{B}_{\rho,m}(\rho,m,x,y)P\Big]\Big{\}},\label{eq25}
\end{align}

with
\begin{align}
\mathcal{A}_{\rho}(\rho,m,x)&=D\partial_{xx}\rho-v \partial_x m,\label{eq26}\\
\mathcal{A}_{m}(\rho,m,x)&=D \partial_{xx}m-v\partial_x\rho-m\Bigg[2\Big(T-\frac{r_3}{4}\Big)+\frac{r_3}{2}\frac{m^3}{\rho^2}\Bigg],\label{eq27}\\
\mathcal{B}_{\rho,\rho}(\rho,m,x,y)&=\Big(-D\partial_{xx}\rho -6v \partial_x m\Big)\delta(y-x),\label{eq28}\\
\mathcal{B}_{m,m}(\rho,m,x,y)&=\Bigg[-D\partial_{xx}\rho -6v \partial_x m + 4\rho\beta\Big(\frac{T+\beta}{\beta}-\frac{m^2}{\rho^2}\Big)\Bigg]\delta(y-x),\label{eq29}\\
\mathcal{B}_{\rho,m}(\rho,m,x,y)&=\Big(-D\partial_{xx}m -6v \partial_x \rho\Big)\delta(y-x),\label{eq30}
\end{align}

and
\begin{equation}
\beta=\frac{r_2}{2}+\frac{r_3}{4}.\label{eq31}
\end{equation}
The corresponding Langevin equation in the Ito sense is:
\begin{align}
\partial_t \rho&=\mathcal{A}_{\rho}(\rho,m,x)+\xi_{\rho}(x,t),\label{eq33}\\
\partial_t m&=\mathcal{A}_{m}(\rho,m,x)+\xi_{m}(x,t),\label{eq34}
\end{align}
where
\begin{align}
\langle\xi_{\rho}(x,t)\xi_{\rho}(y,t^{'})\rangle&=\Delta \mathcal{B}_{\rho,\rho}(\rho,m,x,y)\delta(t-t^{'}),\label{eq35}\\
\langle\xi_{\rho}(x,t)\xi_{m}(y,t^{'})\rangle=\langle\xi_{m}(x,t)\xi_{\rho}(y,t^{'})\rangle&=\Delta \mathcal{B}_{\rho,m}(\rho,m,x,y)\delta(t-t^{'}),\label{eq36}\\
\langle\xi_{m}(x,t)\xi_{m}(y,t^{'})\rangle&=\Delta \mathcal{B}_{m,m}(\rho,m,x,y)\delta(t-t^{'}),\label{eq37}
\end{align}

Since we are interested in the long-wavelength hydrodynamic limit, we can neglect the derivative terms in the stochastic part of the Langevin equations, and set $\mathcal{B}_{\rho,\rho}$ and $\mathcal{B}_{\rho,m}$ equal to zero. We then have our expresion for the stochastic partial differential equation (sPDE) that the system obeys:

\begin{align}
\partial_t \rho&=D\Delta\rho-v \partial_x m,\label{eq38}\\
\partial_t m&=D \Delta m-v\partial_x\rho-m\Bigg[2\Big(T-\frac{r_3}{4}\Big)+\frac{r_3}{2}\frac{m^2}{\rho^2}\Bigg]+2\sqrt{\frac{\beta}{\rho}\Bigg(\frac{T+\beta}{\beta}\rho^2-m^2\Bigg)}\eta,\label{eq39}
\end{align}
where $v=2D\epsilon$, $\beta=(r_2/2)+(r_3/4)$ and $\eta(x,t)$ is a Gaussian white noise that satisfies:
\begin{equation}
\langle \eta(x,t)\eta(y,t^{'})\rangle=\delta(y-x)\delta(t-t^{'}).\label{eq39}
\end{equation}

\section{II. Probability Distribution of $m/\rho$ from WFT}
In this section, we derive the closed form expression for the probability distribution of $m/\rho$, as predicted by the WFT approximation (Eq.($10$) in the main text):
\begin{equation}
P(\rho,m,x,t)=\mathcal{N}(\rho-\bar{\rho},\sigma^2_{\rho})\mathcal{N}(m-\bar{m},\sigma^2_{m}).\label{eq40}
\end{equation}
Given this joint probability distribution for $m$ and $\rho$, $P(m/\rho)$ can be determined as:
\begin{align}
P\Big(\frac{m}{\rho}\Big)&=\int dm^{'} \hspace{5pt}\int d\rho^{'} \hspace{5pt}P(\rho^{'},m^{'})\delta(m/\rho-m^{'}/\rho^{'}),\label{eq41}\\
&=\int d\rho^{'} \hspace{5pt}|\rho^{'}|\hspace{5pt}P\Big(\rho^{'},\frac{m}{\rho}\rho^{'}\Big),\nonumber\\
&=\int d\rho^{'} \hspace{5pt}|\rho^{'}|\hspace{5pt}\mathcal{N}(\rho^{'}-\bar{\rho},\sigma^2_{\rho})\hspace{5pt}\mathcal{N}\Big(\frac{m}{\rho}\rho^{'}-\bar{m},\sigma^2_{m}\Big)\label{eq42}.
\end{align}
After evaluating this integral Eq.(\ref{eq42}), we get:
\begin{align}
P(z)&=\frac{b(z) d(z)}{a^3(z)}\frac{1}{\sqrt{2\pi}\sigma_m\sigma_{\rho}}\Bigg[\Phi\Bigg(\frac{b(z)}{a(z)}\Bigg)-\Phi\Bigg(-\frac{b(z)}{a(z)}\Bigg)\Bigg]\nonumber\\
&+\frac{e^{-c/2}}{a^2(z)\pi\sigma_m\sigma_{\rho}},\label{eq43}
\end{align}
where
\begin{align}
z&=m/\rho,\label{eq44}\\
a(z)&=\sqrt{\frac{z^2}{\sigma_m^2}+\frac{1}{\sigma_{\rho}^2}},\label{eq45}\\
b(z)&=\frac{\bar{m}}{\sigma_m^2}z+\frac{\bar{\rho}}{\sigma_{\rho}^2},\label{eq46}\\
c&=\frac{\bar{m}^2}{\sigma_m^2}+\frac{\bar{\rho}^2}{\sigma_{\rho}^2},\label{eq47}\\
d(z)&=\frac{e^{b^2(z)-ca^2(z)}}{2a^2(z)},\label{eq48}
\end{align}
and
\begin{equation}
\Phi(t)=\int_{-\infty}^{t} du \hspace{5pt}\frac{e^{-u^2/2}}{\sqrt{2\pi}},\label{eq49}
\end{equation}
is the cumulative distribution function for the normal distribution. For zero mean, $\bar{m}=\bar{\rho}=0$, and unit variance, $\sigma^2_{m}=\sigma^2_{\rho}=1$, $P(m/\rho)$ is nothing but the Cauchy distribution,
\begin{equation}\label{eq50}
P(m/\rho)=\frac{1}{\pi\Big(\frac{m^2}{\rho^2}+1\Big)}.
\end{equation}
The general case with non-zero means and variances not equal to one, results in a shifted Cauchy distribution. It is this $P(m/\rho)$ that is plotted as the red solid curve in Fig. $4$(b) of the main text, and  when compared to the probability distribution of $m/\rho$ from exact simulations (blue histograms), shows good agreement.

\end{document}